\documentclass[fleqn]{article}
\usepackage[dvips]{graphics}
\usepackage{epsf}

\newcommand{\bls}[1]{\renewcommand{\baselinestretch}{#1}}
\newcommand{\Abstract}[1]{\vskip 2mm \begin{center}
        \parbox{16.4cm}{\small\noi #1} \end{center}\medskip}
\newcommand{\noi}{\noindent}
\newcommand{\Title}[1]{\noi {\Large #1} \\}
\newcommand{\email}[2]{\footnotetext[#1]{e-mail: #2}}

\renewcommand{\theequation}{\arabic{section}.\arabic{equation}}
\catcode`\@=11 \@addtoreset{equation}{section}\catcode`\@=12
\def\beq#1{\begin{equation}\label{#1}}
\def\eeq{\end{equation}}
\def\bearr{\begin{eqnarray} \lal}
\newcommand{\bear}[1]{\begin{eqnarray}\label{#1}}
\def\ear{\end{eqnarray}}

\def\lal{&&\nqq {}}
\def\nn{\nonumber\\ {}}
\def\al{&\hspace*{-0.5em}}
\def\eql{\al =\al}
\def\yy{\\[5pt] {}}
\def\nqq{\hspace*{-2em}}
\def\dys{\displaystyle}
\def\mst{\mathstrut}

\def\e{{\,\rm e}}
\newcommand{\R}{{\bf R}}
\newcommand{\sh}{\mathop{\rm sh}\nolimits}
\def\sign{\mathop{\rm sign}\nolimits}
\newcommand{\ch}{\mathop{\rm ch}\nolimits}

\bls{0.957}

\begin{document}

\Title  {\uppercase{Simulation of intersecting black brane solutions \\\\
by multi-component anisotropic fluid}}

\noi{\large \bf V.D. Ivashchuk,$^{a,b,1}$ V.N. Melnikov$^{a,b,2}$ and
        A.B. Selivanov$^{a,3}$}

\medskip{\protect
\begin{description}\itemsep -1pt
\item[$^a$]
     {\it Institute of Gravitation and Cosmology, Peoples Friendship
                University of Russia \\
      6 Miklukho-Maklaya St., Moscow 117198, Russia}

\item[$^b$]{\it Centre for Gravitation and Fundamental Metrology, VNIIMS\\
    3-1 M.Ulyanovoy St., Moscow 119313, Russia}

\end{description}}

\Abstract {A family of spherically symmetric solutions
with horizon in the model with
multi-component anisotropic fluid (MCAF) is obtained. The metric of
any solution contains $(n-1)$ Ricci-flat ``internal space''
metrics and for certain equations of state ($p_i = \pm \rho$)
coincides with the metric of intersecting black brane
solution in the model with antisymmetric forms. Examples of
simulation of intersecting $M2$ and $M5$ black branes are
considered. The post-Newtonian parameters $\beta$ and $\gamma$
corresponding to the 4-dimensional section of the metric are
calculated.}



\email 1 {ivas@rgs.phys.msu.su} \email 2
{melnikov@rgs.phys.msu.su} \email 3 {seliv@rgs.phys.msu.su}

\section{Introduction}

Recently, spherically-symmetric $p$-brane  solutions with horizon (see,
e.g., \cite{IMtop} and references  therein) defined on product manifolds
$\R \times M_0 \times \ldots  \times M_n$  cause a wide interest. These
solutions appear  in models with antisymmetric forms and scalar fields.
These  and
 more general $p$-brane cosmological and spherically symmetric
solutions are usually obtained by reduction of the field equations
to the Lagrange equations corresponding to Toda-like systems
\cite{IMJ}. An analogous reduction for models with multi-component
anisotropic fluids was performed earlier in \cite{IM5}. For
cosmological-type models with antisymmetric forms without scalar
fields any $p$-brane is equivalent to an anisotropic fluid with
the equations of state:
\beq{0}
  \hat{p}_i =  - {\hat \rho}  \qquad {\rm or} \qquad {\hat p}_i = {\hat
\rho},
 \eeq
when the manifold $M_i$ belongs or does not belong to the
brane world volume, respectively (here ${\hat p}_i$ is the effective pressure
in
 $M_i$  and ${\hat \rho}$ is the effective density).

In this paper we  find the analogues of intersecting black
brane  solutions in a model with multi-component
anisotropic fluid (MCAF), when certain "orthogonality" relations
on fluid parameters are imposed.
The one-component case was considered
earlier in \cite{IMS}.

The paper is organized as follows. In Section 2 the model is
formulated. In Section 3 general MCAF solutions with horizon corresponding to black-brane-type solutions are
presented. Section 4  deals with certain MCAF analogues of
intersecting black brane
solutions, i.e. $M2$ and $M5$ black brane solutions.
In Section
5 the post-Newtonian parameters for the 4-dimensional section of
the MCAF-black-brane metric are calculated. In Appendix
based on \cite{IMtop, GIM}
the general spherically symmetric solutions with multicomponent anisotropic
fluid are considered and configurations with horizon are singled out.

\section{The model}

In this paper we consider a family of spherically symmetric
solutions to Einstein equations with an anisotropic matter source
\beq{1.1}
        R^M_N - \frac{1}{2}\delta^M_N R = k T^M_N,
\eeq defined on the manifold

\beq{1.2}
\begin{array}{l}
M = {\R}_{*}\times (M_{0}=S^{d_0}) \times (M_1 = {\R})
    \times \ldots \times M_n,\\ \qquad ^{\rm radial
    \phantom{p}}_{\rm variable}\quad^{\rm spherical}_{\rm variables}
            \quad\qquad^{\rm time}
\end{array}
\eeq
with the block-diagonal metrics

\beq{1.2a}
    ds^2= \e^{2\gamma (u)} du^{2}+\sum^{n}_{i=0}
            \e^{2X^i(u)} h^{[i]}_{m_i n_i }dy^{m_i}dy^{n_i }.
\eeq
Here $\R_{*} \subseteq \R$ is an open interval. The manifold
$M_i$ with the metric $h^{[i]}$, $i=1,2,\ldots,n$, is a Ricci-flat
space of dimension $d_{i}$: \beq{1.3}
    R_{m_{i}n_{i}}[h^{[i]}]=0,
\eeq and $h^{[0]}$ is the standard metric on the unit sphere
$S^{d_0}$, so that \beq{1.4}
    R_{m_{0}n_{0}}[h^{[0]}]=(d_0-1)h^{[0]}_{m_{0}n_{0}};
\eeq $u$ is a radial variable, $\kappa$ is the gravitational
constant, $d_1 = 1$ and $h^{[1]} = -dt \otimes dt$.

The energy-momentum tensor is adopted in the following form for
each component of the fluid:
\beq{1.5} ({T^{(s)}}^{M}_{N})= {\rm
diag} (-{\hat{\rho}^{(s)}}, {\hat p}_{0}^{(s)}
    \delta^{m_{0}}_{k_{0}}, {\hat p}_{1}^{(s)}
    \delta^{m_{1}}_{k_{1}},\ldots , {\hat p}_n^{(s)}
    \delta^{m_{n}}_{k_{n}}),
\eeq
where $\hat{\rho}^{(s)}$ and $\hat p_{i}^{(s)}$ are the effective
density and pressures respectively, depending on the radial
variable $u$.

We assume that the following "conservation laws"
\beq{5.0}
\nabla_{M}T^{(s)M}_{\ N}=0 \eeq
are valid for all components.

We also impose the following equations of state
\beq{1.7}
    {\hat p}_i^{(s)}=\left(1-\frac{2U_i^{(s)}}{d_i}\right){\hat{\rho}^{(s)}},
\eeq
where $U_i^{(s)}$ are constants, $i= 0,1,\ldots,n$.

The physical density and pressures are related to the effective
ones (with ``hats'') by the formulae
  \beq{1.7a}
        \rho^{(s)} = - {\hat p}_1^{(s)}, \quad p_u^{(s)} = - \hat{\rho}^{(s)},
        \quad p_i^{(s)} = \hat{p}_i^{(s)} \quad (i \neq 1).
   \eeq

In what follows we put $\kappa =1$ for simplicity.

\section{Spherically symmetric solutions with horizon}

We will make the following assumptions: \beq{2.1}
\begin{array}{l}
    1^{o}.\quad U^{(s)}_0 = 0 \
    \Leftrightarrow \ \hat p^{(s)}_0 = \hat\rho^{(s)} ,\yy
    2^o.\quad U^{(s)}_1 = 1 \
    \Leftrightarrow \  \hat p^{(s)}_1 = -\hat\rho^{(s)} ,\yy
    3^o.\quad (U^{(s)},U^{(s)})  = U^{(s)}_i G^{ij}U^{(s)}_j > 0,
    \quad (U^{(s)},U^{(l)}) = 0, \quad s \neq l ,
\end{array}
\eeq where

\beq{2.2a}
    G^{ij}=\frac{\delta^{ij}}{d_i} + \frac{1}{2-D},
\eeq are components of the matrix inverse to the matrix of the
minisuperspace metric \cite{IMZ}

\beq{2.2}
    (G_{ij}) = (d_i \delta_{ij} - d_i d_j),
\eeq and $D=1+\sum\limits_{i=0}^n {d_i}$ is the total dimension.

The orthogonality condition $3^o$ is an integrability condition
(see Appendix). The conditions $1^o$ and $2^o$ in p-brane terms
mean that brane "lives" in a time manifold $M_1$ and does not
"live" in ${\R}_{*}\times M_{0}$. The assumptions $1^o$ and $2^o$
are natural ones from the point of view of state equations
(\ref{1.7}), so we can rewrite the energy-momentum tensor
(\ref{1.5}) as following:

\beq{2.1a}
    ({T^{(s)}}^{M}_{N})= {\rm diag} (-\rho^{(s)},\
    \rho^{(s)} \delta^{m_0}_{k_0},\
    - \rho^{(s)} \delta^{m_1}_{k_1},\
    p_2^{(s)}\delta^{m_2}_{k_2},\ \ldots ,
    p_n^{(s)}\delta^{m_n}_{k_n}).
\eeq

Under the conditions (\ref{1.7}) and (\ref{2.1}) we have obtained
the following black-hole solutions to the Einstein equations
(\ref{1.1}):

\bear{12} \lal
    ds^{2} = J_{0}\left( \dys\frac{\mst
    dr^{2}}{1-{2\mu}/{r^{d}}} + r^{2} d \Omega^2_{d_0} \right) -
        J_1\left(1-\frac{2\mu}{r^{d}}\right)dt^{2}
    + \sum_{i=2}^{n} J_{i} h^{[i ]}_{m_{i}n_{i}} dy^{m_{i}}dy^{n_{i}},
\\  \lal \label{13}
    \rho^{(s)}= - \dys\frac{\mst A_s}{H_s^2 J_0 r^{2d_0}} ,
    \qquad A_{s} = - \frac12 \ \nu_s^2 d^2 P_{s} (P_{s} +2\mu),
\ear
 which may be verified from \cite{IM5} and by analogy with the
 $p$-brane solution
\cite{IMJ}. For direct derivation of the solution see Appendix. Here
$d=d_0-1$, \beq{2.sp}
    d \Omega_{d_0}^2= h^{[0]}_{m_{0}n_{0}} dy^{m_{0}}dy^{n_{0}}
\eeq
is the spherical element,

\beq{2.3}
    J_{i} = \prod_{s =1}^m H_s^{-2\nu_s^{2}U^{(s) i }},
    \qquad H_s =1+{P_{s}}/{r^{d}};
\eeq
$P_{s} >0$, $\mu >0$ are integration constants and
\bearr\label{2.4}
    U^{(s)  i} = G^{ij}U^{(s)}_{j}  = \frac{U^{(s)}_i}{d_i} + \frac{1}{2-D}
                \sum_{j=0}^{n}U^{(s)}_j ,
\\ \lal \label{2.4a}
        \nu_s = ({U^{(s)}},{U^{(s)}})^{-1/2}.
\ear

\section{Simulation of intersecting black branes}

The solution from the previous section for MCAF allows to simulate the
intersecting black brane solutions \cite{IMtop}
in the model with antisymmetric forms without scalar fields.
In this case the parameters $U^{(s)}_i$ have  the following
form:
\beq{3.1a}
\begin{array}{ccccccr}
 U^{(s)}_i & = & d_i , & p_i^{(s)} & = & -\rho^{(s)} , &  i\in I_{(s)};\\
    &  & 0 , &  & & \rho^{(s)} ,  &  i \notin I_{(s)}.
\end{array}
\eeq
Here
 $I_{(s)} = \{ i_1, \ldots,  i_k \} \in \{1, \ldots n \}$
is the index set \cite{IMtop} corresponding to brane submanifold
$M_{i_1}  \times \ldots \times M_{i_k}$.

The orthogonality constraints  $3^o$  (\ref{2.1}) lead us to  the following
dimension of intersection of brane submanifolds \cite{IMtop}:

\beq{3.1b} d_{I_{(s)}\cap
I_{(l)}}=\frac{d_{I_{(s)}}d_{I_{(l)}}}{D-2}, \eeq where
$d_{I_{(s)}}$ and $d_{I_{(l)}}$ are dimensions of $p$-brane
world-volumes, $s, l = 1,\ldots ,m$, $s \neq l$.

Due to relations (\ref{3.1a}) and  $1^o$,  (\ref{2.1}) we can rewrite
(\ref{13}) as follows:

\beq{3.1d}
    \rho^{(s)}= - \dys\frac{\mst A_s} {H_s^2 \prod\limits_{l=1}^m
    H_l^{2/(D-d_{I_{(l)}}-2)} r^{2d_0}},
\eeq
and investigate the behavior of the density as a radial function.
For the single fluid the density is regular and positive at zero
when the parameter $d$ (see the previous section) is equal to
$d^*=D-d_{I_{(1)}}-2$. In this case the brane submanifold fills
the total manifold (\ref{1.2}) except $R_* \times S^{d_0}$.When $d
< d^*$ the density is infinite at zero.

For multi-component fluid all densities are finite at $r=0$, if
(and only if)

\beq{3.1e}
\sum_{s=1}^{m}\frac{1}{D-d_{I_s}-2} \geq \frac{1}{d}.
\eeq
Moreover, all $\rho^{(s)}(0)>0$ when the equality
in (\ref{3.1e}) takes place.

\begin{figure}
\centering\includegraphics{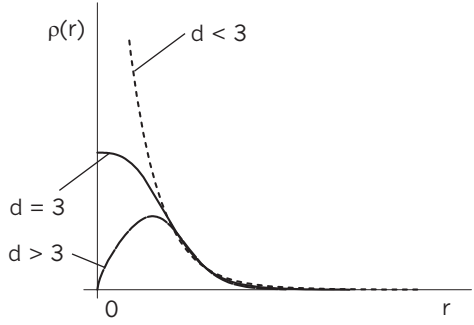}
  \caption{The variants of behavior of $\rho(r)$ for $M2 \cap M2$ intersection}
\end{figure}

As an example we consider simulation by MCAF of intersecting $M2
\cap M5$, $M2 \cap M2$, $M5 \cap M5$ configurations in $D = 11$
supergravity. The metric for all cases reads:

\bearr\label{3.3c}
ds^2 =  J_0\Biggl[\dys\frac{\mst
        dr^2 }{1- 2\mu/r} + r^2 d \Omega^2_{d_0} -
        {(H_{(I)})}^{-1}{(H_{(II)})}^{-1}\biggl\{
        \left(1-\frac{2\mu}{r}\right) dt^2
        + h^{[2]}_{m_2 n_2}dy^{m_2}dy^{n_2} \biggr\} \nn &&
        + H_{(I)}^{-1} h^{[3]}_{m_{3} n_{3}} dy^{m_3}dy^{n_3}
        + H_{(II)}^{-1} h^{[4]}_{m_{4} n_{4}} dy^{m_4}dy^{n_4}
        + h^{[5]}_{m_{5}n_{5}} dy^{m_{5}}dy^{n_{5}} \Biggr],
\ear

where we can express the factor $J_0=H_{(I)}^{2/(D-d_{I_{(I)}}-2)}
H_{(II)}^{2/(D-d_{I_{(II)}}-2)}$; the first brane world-volume is
$M_1 \times M_2 \times M_3$, the second one is $M_1 \times M_2
\times M_4$.

$a).$ For MCAF, corresponding to intersecting of $M2$ (with index
$I$ in (\ref{3.3c})) and $M5$ (with index $II$) branes  the
dimensions are following $d_1 =d_2 =d_3 =1,\ d_4 =4$ and
$J_0=H_{(I)}^{1/3}H_{(II)}^{2/3}$.

The densities $\rho^{( I )},\ \rho^{( II )}$  are infinite at zero
when $d=1$ and for $d=2$ they are finite:
$\rho^{( I )}(0)=(P_{I}+2\mu)/H_{I}^{4/3} H_{II}^{2/3}$, $\rho^{( II )}(0)=
(P_{II}+2\mu)/H_{II}^{5/3} H_{I}^{1/3}$. It is interesting to note
that in the extremal limit $\mu \rightarrow 0$ $\rho^{( I
)}(0)=\rho^{( II )}(0)$.

$b).$ For MCAF equivalent to two electrical $M2$ branes
intersecting on the time manifold  we get $d_3 = d_4 = 2,\ d_2 =
0$. Here $J_0=H_{(I)}^{1/3}H_{(II)}^{1/3}$.

The variants of behavior of the densities are presented on Figure
1. When $d=3$ both functions are regular and positive at zero (the
middle branch).

$c).$ For two $M5$ branes the dimension of intersection is 4 and
$d_0 =d_3 = d_4 = 2,\ d_2 = 3,\ d_5 = 0$ and
$J_0=H_{(I)}^{2/3}H_{(II)}^{2/3}$. The only possibility here is
$d=1$ and the fluid densities are infinite at zero.

\section{Physical parameters}

\subsection{Gravitational mass and post-Newtonian parameters}

Here for simplicity we put $d_0  =2 \ (d =1)$. Consider the 4-dimensional
space-time section of the metric (\ref{12}). Introducing a new
radial variable by the relation

\beq{3.7}
    r = R \left(1 + \frac{\mu}{2 R}\right)^2,
\eeq we rewrite the 4-section in the following form:
\bear{3.8}\lal
    ds_{(4)}^2 = g^{(4)}_{\mu \mu'} dx^{\mu} dx^{\mu'} =
    \left(\prod_{s=1}^m H_s^{- 2\nu_s^2 U^{(s)0}}\right)
    \left[ - \left(\frac{1-{\mu}/{2 R}}{1+{\mu}/{2 R}}\right)^2
    \left(\prod_{s=1}^m H_s^{- 2\nu_s^2 }\right)dt^2 +
\left( 1+\frac{\mu}{2 R} \right)^4 \delta_{ij} dx^i dx^j \right],
\ear $i,j = 1,2,3$. Here $R^2 = \delta_{ij} x^i x^j$.

The post-Newtonian (Eddington) parameters are defined by the
well-known relations \bear{3.9}
    g^{(4)}_{00} \eql - (1-2V+2\beta V^2) + O (V^3), \\
\label{3.10}
    g^{(4)}_{ij} \eql \delta_{ij} (1 + 2 \gamma V) + O(V^2),
\ear $i,j = 1,2,3$. Here $V={GM}/{R}$ is the Newtonian potential,
$M$ is the gravitational mass and $G$ is the gravitational
constant. From (\ref{3.8})-(\ref{3.10}) we obtain:
\beq{3.12}
    GM = \mu + \sum_{s=1}^m \nu_s^2 P_{s} (1 +U^{(s)0}),
\eeq
and
\bear{3.13}
    \beta - 1 \eql \frac{1}{2 (GM)^2}\sum_{s=1}^m \ \nu_s^2
    P_{s} (P_{s} + 2 \mu) (1 + U^{(s) 0}), \\
\label{3.13a}
    \gamma - 1 \eql  - \frac{1}{GM}\sum_{s=1}^m \
    \nu_s^2 P_{s} (1 + 2U^{(s) 0}).
\ear

For fixed vector $U^{(s)}$ the parameter $\beta -1$ is proportional
to the ratio of two physical parameters:  the anisotropic fluid
density parameter $A_{s}$ (see (\ref{5.78a})), and the
gravitational radius squared $(GM)^2$.

\subsection{The Hawking temperature}

The Hawking temperature of a black hole may be calculated using
the relation from \cite{York} and has the following form:
\beq{5.79}
    T_H = \frac{d}{4 \pi (2 \mu)^{1/d}} \prod_{s=1}^m
    \left(\frac{2 \mu}{2 \mu + P_{s}}\right)^{\nu_s^2}.
\eeq

\section{Conclusions}

Here we have obtained a family of spherically symmetric
solutions with horizon in the model
with multi-component anisotropic fluid with the equations of state
(\ref{1.7}) and the conditions (\ref{2.1}) imposed.  The metric of
any solution  contains $(n -1)$ Ricci-flat ``internal'' space
metrics. For certain equations of state (with $p_i = \pm \rho$)
the metric of the solution  may coincide with the
metric of intersecting black branes
(in a model with antisymmetric forms
without dilatons). Here the examples of simulating
of intersecting  $M 2$ and $M 5$ black branes in $D=11$
supergravity are considered.

We have also calculated the post-Newtonian parameters $\beta$ and
$\gamma$ corresponding to the 4-dimensional section of the metric.
The parameter $\beta -1$ is written in terms of ratios of the
physical parameters: the anisotropic fluid parameter $|A_{s}|$ and
the gravitational radius squared $(GM)^2$. An open problem
is to generalize the formalism to the case when dilaton
scalar fields are added into consideration.

{\bf Acknowlegments}

This work was supported in part by the Russian Ministry of Science
and Technology, Russian Foundation for Basic Research (Grant
01-02-17312), project SEE and DFG project
(436 RUS 113/678/0-1(R)).

V.D.I. and V.N.M. thank Prof. Dr. H. Dehnen
and his colleagues at the University of Konstanz for their hospitality.


\renewcommand{\theequation}{\Alph{subsection}.\arabic{equation}}
\renewcommand{\thesection}{}
\renewcommand{\thesubsection}{\Alph{subsection}}
\setcounter{section}{0}

\section{Appendix}

\subsection{Lagrange representation}

The "conservation law" equation (\ref{5.0}) may be written,
due to relations (\ref{1.2a}) and (\ref{1.5}) in the following form:

\beq{5.7} \dot{\hat{\rho}}^{(s)} +\sum_{i=0}^n
d_i\dot{X^i}({\hat{\rho}}^{(s)} +{\hat p}^{(s)}_i )=0. \eeq Using
the equation of state (\ref{1.7}) we get

\beq{5.7a} {\hat{\rho}}^{(s)}= - A_{s} e^{2U^{(s)}_i
X^{i}-2\gamma_{0}}, \eeq
where $\gamma_0(X)= \sum\limits_{i=0}^{n} d_{i}X^{i}$, and $A_{s}$
are constants.

The Einstein equations (\ref{1.1})  with the relations
(\ref{1.7}) and (\ref{5.7a}) imposed are equivalent to the
Lagrange equations for the Lagrangian


\beq{} L = \frac
{1}{2}e^{-\gamma+\gamma_0(X)}G_{ij}\dot{X}^{i}\dot{X}^{j}
-e^{\gamma-\gamma_0(X)}V, \eeq
where

\beq{5.32n}
V= \frac{1}{2} d_0 (d_0 -1) e^{2U^{(0)}_i X^i} +
\sum_{s= 1}^m A_{s} e^{2 U^{(s)}_{i}X^{i}} = \sum_{s = 0}^m A_{s}
e^{2 U^{(s)}_{i}X^{i}}, \eeq
is the potential and the components
of the minisupermetric $G_{ij}$ are defined in (\ref{2.2}).
\beq{5.8}
U^{(0)}_i X^i = -X^0 + \gamma_0(X), \qquad U^{(0)}_i  =
- \delta^0_i + d_i, \qquad A_{0} = \frac{1}{2} d_0 (d_0 -1),
\eeq
$i = 0, \ldots, n$.

For $\gamma=\gamma_0(X)$, i.e. when the harmonic time gauge
is considered, we get the set of Lagrange equations
for the Lagrangian
\beq{5.31n}
L=\frac12G_{ij} \dot X^i \dot X^j-V,
\eeq
with the zero-energy constraint imposed
\beq{5.33n}
E=\frac12G_{ij} \dot X^i \dot X^j + V =0.
\eeq

It follows from the restriction
$U^{(s)}_0 = 0$ that \beq{5.43a}
(U^{(0)},{U^{(s)}})  \equiv U^{(0)}_i G^{ij}U^{(s)}_j = 0.
\eeq

Indeed, the contravariant components $U^{(0)i}=
G^{ij} U^{(0)}_j$ are the following ones
\beq{5.43b}
U^{(0)i}=-\frac{\delta_0^i}{d_0}.
\eeq

Then we get $(U^{(0)},U^{(s)})  = U^{(0)i} U^{(s)}_i = -
U^{(s)}_0/d_0 =0$. In what follows we also use the formula

\beq{5.43c}
\frac{1}{\nu_0^2} = (U^{(0)},U^{(0)})   = \frac{1}{d_0} - 1 < 0,
\eeq
for $d_0 >1$.

In what follows we will make the following assumption on indices:
   $s = 1, \ldots, m$ and $\alpha= 0, \ldots, m$.

\addtocounter{section}{1} \setcounter{equation}{0}
\subsection{General spherically symmetric and cosmological-type
solutions}

When the orthogonality relations  (\ref{5.43a}) and
$3^o$ of (\ref{2.1}) are satisfied the Euler-Lagrange equations for the
Lagrangian (\ref{5.31n}) with the potential (\ref{5.32n}) have the
following solutions (see relations from \cite{GIM} adopted for our
case):

\beq{5.34n} X^i(u)= -
\sum_{\alpha=0}^m\frac{U^{(\alpha)i}}{(U^{(\alpha)},U^{(\alpha)})}\ln
|f_\alpha(u-u_\alpha)| + c^i u + \bar{c}^i, \eeq
where $u_\alpha$ ($\alpha= 0, \ldots, m$) are integration
constants; and vectors $c=(c^i)$ and $\bar c=(\bar c^i)$ are
orthogonal to the $U^{(\alpha)}=(U^{(\alpha)i})$, i.e. they
satisfy the linear constraint relations

\bear{5.47n}
U^{(0)}(c)= U^{(0)}_i c^i = -c^0+\sum_{j=0}^nd_jc^j=0, \\
\label{5.48n} U^{(0)}(\bar c)= U^{(0)}_i \bar c^i = -\bar
c^0+\sum_{j=0}^nd_j\bar c^j=0, \\
\label{5.49n} U^{(s)}(c)= U^{(s)}_i c^i=0,\\
\label{5.50n} U^{(s)}(\bar c)=  U^{(s)}_i\bar c^i=0. \ear

Here
\beq{A.7}
\begin{array}{rlll}
f_\alpha(\tau)=
 & R_\alpha
 \dys\frac{\mst\sh(\dys\sqrt{\mst C_\alpha}\tau)}{\dys\sqrt{\mst C_\alpha}},
 & C_\alpha\neq 0,&
\eta_\alpha = +1 , \\  & R_\alpha \dys\frac{\mst\ch(\dys\sqrt{\mst
C_\alpha}\tau)}{\dys\sqrt{\mst C_\alpha}},
& C_\alpha>0, & \eta_\alpha = -1 , \\\\
 & R_\alpha \tau,& C_\alpha=0,& \eta_\alpha = +1 ,
\end{array}
\eeq
$R_{\alpha} =\sqrt{2|A_{\alpha}/\nu_{\alpha}^2|}$, $\eta_{\alpha}
= - \sign (A_{\alpha}/\nu_{\alpha}^2)$; and parameters
$\nu_{\alpha}$ are defined in (\ref{2.4a}) and (\ref{5.43c}),
$\alpha= 0, \ldots, m$.

The zero-energy constraint, corresponding to the solution (\ref{5.34n})
reads

\beq{A.17} E = \frac12 \sum_{\alpha = 0}^m
\frac{C_\alpha}{(U^{(\alpha)} , U^{(\alpha)})} + \frac12
G_{ij}c^ic^j=0 . \eeq

From (\ref{5.34n}) we get the following relation for the metric
(see  also (\ref{2.2}), (\ref{5.43b}) and (\ref{5.43c}))

\beq{5.63n}
\begin{array}{l}
g=\e^{2c^0u+2\bar c^0} \ \left(\prod\limits_{\alpha = 0}^m
f_\alpha^{\dys^{\mst -2 \nu^2 U^{(\alpha) 0}}}\right) \biggl\{\
du\otimes du + f_0^2 \ h^{[0]} \biggr\}+ \sum\limits_{i\ne0}
 \ \e^{2c^iu+2\bar c^i}
 \ \left(\prod\limits_{\alpha = 0}^m f_\alpha^{\dys^{\mst -2 \nu^2
U^{(\alpha) i}}} \right) \ h^{[i]},
\end{array}
\eeq where $f_\alpha = f_\alpha (u-u_\alpha)$ (here we use the
relations $d_i U^i + \frac{U_0}{d_0}= U^0$ and (\ref{5.43c})).

{\bf Solutions with horizon.} For integration constants we put
\bear{5.67}
\bar{c}^i  & = &  0,\\
c^i & = & \bar{\mu}  \sum_{\alpha = 0}^m
\frac{U^{{(\alpha)}i}}{(U^{(\alpha)},U^{(\alpha)})} -
\bar{\mu} \delta^i_1, \\
       \label{5.68}
C_\alpha  & = &  \bar{\mu}^2, \ear where $\bar{\mu} > 0$,
$\alpha=0, \ldots,m$.

We also introduce new radial variable $r = r(u)$ by relations

\beq{5.69} \exp( - 2\bar{\mu} u) = 1 - \frac{2\mu}{r^d},  \quad
\mu = \bar{\mu}/ d >0, \quad  d = d_0 -1, \eeq

and put $u_0 = 0$; $u_s < 0$, $A_{s}  < 0, \ s=1,\ldots,m$,

\beq{5.70} \frac{\sqrt{2|A_{s}|}}{\bar{\mu} \nu_s} \sh \beta_s =1,
\quad \beta_s \equiv \bar{\mu}| u_s|. \eeq

Now the parameter $P_{s}$ may be introduced ($P_{s} > 0$) by the
following relation:

\beq{5.78} \frac{\mu}{\sh \beta_s} = P_{s} e^{\beta_s} = \sqrt{
P_{s} ( P_{s} +2\mu)}, \eeq and, hence,

\beq{5.78a} -A_{s} = \frac12 \nu_s^2 d^2 P_{s} (P_{s} +2\mu), \eeq

see (\ref{5.7a}). The relations of the Appendix imply the formulae
(\ref{12}), (\ref{13}) for the solution from Section 3.

\newpage
\begin{center} {\bf List of captions for illustrations}
\end{center}

Figure 1. The variants of behavior of $\rho(r)$ for $M2 \cap M2$ intersection.

\end{document}